\documentclass[12pt]{article}
\usepackage{amsbsy}
\usepackage{amsfonts}
\usepackage{amsmath}
\usepackage{amssymb}
\usepackage{apacite}
\usepackage{setspace}
\usepackage{graphicx}
\usepackage{bm,multicol}
\usepackage[english]{babel}
\usepackage{natbib}
\usepackage[T1]{fontenc}
\usepackage[utf8]{inputenc}
\usepackage{authblk}
\usepackage{xcolor}
\usepackage{geometry}
\newcommand{\newc}{\newcommand}
\newc{\N}{\mbox{N}}
\newc{\1}{\bf{1}}

\begin{document}
\title{Comment on ``Safe Testing'' by Gr\"{u}nwald, de Heide, and Koolen}

\author{J. Mulder, Tilburg University}

\maketitle

The potential of $e$-values \citep{safetesting} to unite different schools of statistics and allowing optional stopping in a `safe' manner is a major accomplishment of the authors. The theory implies a change in focus from classical error rates to the expected evidence against the null under the respective hypotheses.

Under $\mathcal{H}_0$, it is required that the expected evidence against the null should not exceed 1, i.e., $\textbf{E}_{\mathcal{H}_0}[E]\le 1$, as in their formula (1). Clearly, using the reciprocal of $p$ values as a measure of evidence severely violates this criterion as $\textbf{E}_{\mathcal{H}_0}[1/p]=\infty$. Thus, we see (via a new route) that the routine use of $p$ values in scientific practice can seriously harm the trustworthiness of research findings.

Under $\mathcal{H}_1$, Bayes factors using a right Haar prior on the scale behave as $e$-values with an optimal growth rate. More (nontrivial) $e$-values exist as stated by the authors. Here I consider the fractional Bayes factor \citep[FBF;][]{OHagan:1995} which are easy to compute and easy to apply as no proper priors need to be formulated.

\begin{figure}[t]
\begin{center}
\includegraphics[height=7cm]{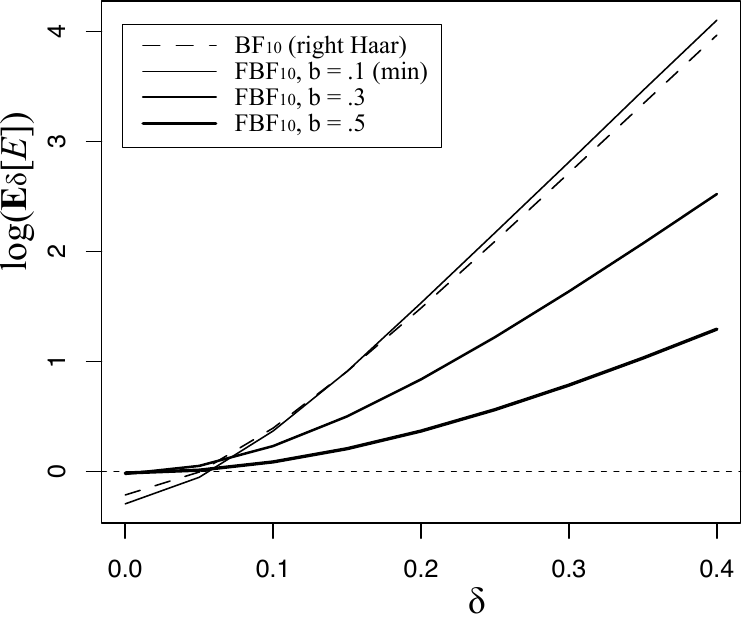}
\caption{Logarithm of the expected evidence $E$ for $\mathcal{H}_1:\delta\not =0$ against $\mathcal{H}_0:\delta=0$ using FBFs with different fractions $b$ and the Bayes factor with the right Haar prior in case $n=20$ and varying standardized effects $\delta$. The expected values were computed numerically using 1e7 randomly generated data. The horizontal dotted line denotes requirement (1) on a log-scale.}
\label{FBF_lines}
\end{center}
\end{figure}

In the FBF, a (minimal) fraction of the data, $b$, is used to update an improper prior yielding a proper `data adapted prior' while the remaining (maximal) fraction is used for computing the evidence \citep{Gilks:1995,Mulder:2014b}. For a simple $t$-test, the FBF of $\mathcal{H}_1$ against $\mathcal{H}_0$ is also scale-invariant and given by $\frac{\Gamma(\frac{nb}{2})\Gamma(\frac{n-1}{2})}{\Gamma(\frac{nb-1}{2})\Gamma(\frac{n}{2})}(1+(n-1)^{-1}t^2)^{-n(b-1)/2}$, where $t$ is the regular $t$-statistic. For $n=20$, Figure \ref{FBF_lines} shows the expected evidence based on a FBF with a minimal fraction of $\frac{2}{n}=.1$, computed with `BFpack' in R \citep{mulder2021bfpack}, and larger fractions, and using a Bayes factor based on a right Haar prior on the scale \citep{Rouder:2009} computed with the `BayesFactor' package \citep{morey2015package}, all on a logarithmic scale. The figure suggests that FBFs (i) are `safe' as requirement (1) under $\mathcal{H}_0$ is satisfied, and (ii) behave competitively to a Bayes factor with optimal growth rate under $\mathcal{H}_1$ when using a minimal fraction, with even higher growth rates for larger effects. Formal proofs are left for future work. Finally note that in sequential testing scenarios, FBFs need to be recomputed on the entire (combined) data in order to be coherent. This can simply be done by storing the sufficient statistics of the data batches.

\bibliographystyle{apacite}
\bibliography{refs_mulder}

\end{document}